# Tuning the superconducting state in the magnetic MoSr$_2$YCu$_2$O$_8$ materials

I. Felner and E. Galstyan

Racah Institute of Physics, The Hebrew University, Jerusalem, 91904, Israel.

**Abstract**

The properties of the MoSr$_2$YCu$_2$O$_8$ materials are found to systematically change with the oxygen concentration determined by the sintering and annealing conditions. The as-prepared (asp) sample is anti-ferromagnetically (AFM) ordered at 16 K. The magnetic features are related to the Mo$^{5+}$ sublattice. Annealing under an oxygen atmosphere induces superconductivity (SC) in the Cu-O planes, and the T$_C$ and the shielding fraction values obtained depend strongly on the oxygen concentration. Annealing the asp material at 1030°C under ambient oxygen atmosphere yields T$_C$ of 18 K, whereas further annealing at 650° C under 92 atm. of oxygen, shifts T$_C$ to 32 K. The AFM ordering, which coexists with SC state through effectively decoupled subsystems, is not affected by the presence or absence of the SC state. In all samples studied, T$_N$ < T$_C$. This behavior resembles most of the inter-metallic magneto-superconductors, but is in sharp contrast to the iso-structural RuSr$_2$GdCu$_2$O$_8$ system where T$_N$ > T$_C$.

**Introduction**

The coexistence of weak ferromagnetism (W-FM) and superconductivity (SC) was discovered a few years ago in RuSr$_2$Ln$_{1.5}$Ce$_{0.5}$Cu$_2$O$_{10}$ (Ln=Eu and Gd, Ru-1222) layered cuprate systems[1-2], and more recently in RuSr$_2$GdCu$_2$O$_8$ (Ru-1212) [3]. Both Ru-based layered cuprate systems evolve from the YBa$_2$Cu$_3$O$_x$ structure where, the Ru ions reside in the Cu (1) site, and only one distinct Cu site (corresponding to Cu (2) in YBa$_2$Cu$_3$O$_x$) with fivefold pyramidal coordination, exists. The SC charge carriers originate from the CuO$_2$ planes and the W-FM state is confined to the Ru layers. In both systems, the magnetic order does not vanish and remains unchanged when SC sets in. Ru-1212 is SC around T$_C$= 32 K and displays a magnetic transition and at T$_M$ =135 K, thus T$_M$ /T$_c$~ 4. The Ru-1222 materials order magnetically at T$_M$= 125-180 K and show bulk SC below T$_C$ = 32-50 K depending on the oxygen and/or Ce concentrations and on sample preparation[4]. The magnetic state of the Ru sublattice is not affected by the presence or absence the SC State, indicating that the two states are practically decoupled.

Partial substitution of Mo for Cu(1) in the tetragonal $YBaSrCu_3O_x$ sample shows that the oxygen content is increased and as a result $T_C$ (81 K for un-doped material ) is shifted to 86 K for 3 at % of Mo[5]. Recently we have noticed that the Ru ions in both Ru-1222 and Ru-1212 systems, can be replaced completely by Mo ions. The $MoSr_2R_{1.5}Ce_{0.5}Cu_2O_{10}$ (Mo-1222R) and $MoSr_2RCu_2O_8$ (Mo-1212R) systems can be obtained with most of the rare-earth (R) elements (Pr-Yb and Y) as nearly single-phase materials[6-7]. We have shown that in Mo-1212R, the magnetic and/or the SC states are determined mainly by the ionic radii of R [7]. For the light R ions (Pr and Nd) the materials are paramagnetic (PM) down to 4 K, whereas in the intermediate range (Sm-Tb), the Mo sublattice orders AFM at $T_N$, ranging from 11 to 24 K. For the heavy R ions (Ho, Er, Tm and Y), coexistence of SC and AFM is observed. SC appears at $T_C$ in the range 19-27 K and AFM sets in at $T_N < T_C$. This is in contrast to the coexistence of SC and W-FM, observed in both Ru-1212 and Ru-1222 systems in which $T_M > T_C$.

Since Y ions are non-magnetic, the study of $MoSr_2YCu_2O_8$ (Mo-1212Y) permits an easier direct interpretation of the SC and the intrinsic Mo AFM states. In this paper, we cover the physical properties of the system only. It is shown that sample prepared at ambient pressure is AFM at $T_N$ =16 K. The magnetic state is not affected by the presence or absence of the SC state. On the other hand, the SC state at $T_C$ 18-32 K depends strongly on the oxygen concentration determined by the sintering and annealing conditions. Small differences of the annealing conditions are very important for the overall oxygen content and affect significantly the $T_C$ and shielding fraction (SF) of the material. This finding is reminiscent of the typical behavior of the development of the SC in the well-known $YBa_2Cu_3O_x$ system.

## Experimental details

Ceramic samples with nominal composition $MoSr_2YCu_2O_8$ were prepared by a solid-state reaction technique. Prescribed amounts of $Y_2O_3$, $SrCO_3$, Mo, and CuO were mixed and pressed into pellets and preheated at 750° C for 1 day. The products were cooled, reground and sintered at elevated temperatures under various conditions. DC zero-field-cooled (ZFC) and field-cooled (FC) magnetic measurements in the range of 5-300 K were performed in a commercial (Quantum Design) super-conducting quantum interference device (SQUID) magnetometer. The resistance was measured by a standard four contact probe and the ac susceptibility was measured by a home-made probe, with an excitation frequency and amplitude of 733 Hz and 30 mOe respectively, both inserted in the SQUID magnetometer. The microstructure and the phase integrity of the materials were investigated by QUANTA (Fri Company) scanning electron microscopy (SEM) and by a Genesis energy depressive x-ray analysis (EDAX) device attached to the SEM.

**Experimental results and discussion**

The $MoSr_2YCu_2O_8$ samples were sintered and annealed several times under various conditions (see below), checked for phase purity and for the magnetic and SC properties. All studied materials were nearly single phase when sintered at least at 1030° C. At lower temperatures multi-phase materials are formed. The as-prepared (asp) non-SC sample was sintered at 1030° C under ambient pressure. Sample (B) was first sintered at 950° C for 24 h and then annealed at 1030° C for 24 h under flowing oxygen. This sample shows a small fraction of the SC state at $T_C$=18 K. Further annealing at 1030° C under flowing oxygen for 24 h (sample C) increases the SC fraction. Sample (D) was obtained by further annealing under flowing oxygen of sample (C) at 600° C for 48 h, and sample (E) is another piece of sample (C) which was annealed for 20 h under high oxygen pressure (92 atm.) at 650° C.

Powder X-ray diffraction (XRD) measurements indicate that all samples are nearly single-phase (~96%) materials and confirmed the tetragonal structure (space group P4/mmm). Due to incomplete reaction, all XRD patterns left a few minor reflections, some of them belonging to the $SrMoO_4$ phase (Fig. 1). All attempts to completely get rid of them were unsuccessful. Least squares fits of the XRD patterns yield within the limit of uncertainty the same lattice parameters: $a$= 3.811(1) Å and $c$= 11.52(2) Å, for all samples studied. In the next sections, assessments of the experimental results performed on the various samples are described.

**(a) The magnetic asp-$MoSr_2YCu_2O_8$**

Figure 2 shows the ZFC and FC plots of the asp-Mo-1212Y sample measured at 40 Oe and 5 kOe. Apart from the magnetic irreversibility, the main effect to be seen is the peak obtained around 11(1) K in both branches indicating an antiferromagnetic (AFM) ordering. The magnetic transition at $T_N$ =16(1) K is defined as the merging temperature of the two curves ($T_{irr}$) when measured at low applied fields. $T_{irr}$ is field dependent, and shifts with the applied field to lower temperatures, (inset Fig.2). Note, that at 5 kOe, $T_{irr}$ is below the peak position. Fig. 1 shows the existence of the Pauli-paramagnetic $SrMoO_4$ phase[8] as an impurity phase. However, its small temperature independent susceptibility is negligible. It is also possible that small amount (not detected by XRD) of the W-FM $Y_2CuO_4$ ($T_N$ 260 K) is also present. However this material reveals the magnetism of the $CuO_2$ sheets only when it is synthesized under high pressure[9]. Therefore, we may associate the AFM features shown in Fig. 2 with the Mo sublattice of the Mo-1212Y structure.

Isothermal M(H) measurements on the asp sample up to 50 kOe at various temperatures, have been carried out and the virgin curve obtained (at 5 K) is presented in Fig. 3. The linear M(H) curve up to about 15 kOe is typical of an AFM substance. Above this field the slop changes slightly indicating a tiny canting of the Mo moments. This canting produces a small hysteresis loop (Fig. 3 inset) with a small remnant moment (2.7 emu/mol) and a coercive field of ~ 100 Oe. The surface morphology of the asp material (Fig. 4a) detected by SEM, shows that this sample has a granular structure with a typical grain size of 2-3 μm. However the grains are not well defined and isolated. We also observed a few separate spherical grains belong to the $SrMoO_4$ phase. The EDAX analysis confirms the initial stoichiometric composition of Mo:Sr:Y and Cu without any deficiency in the Mo content. The composition was found to be uniform from grain to grain.

### (b) The SC and magnetic $MoSr_2YCu_2O_8$

For samples annealed under oxygen flowing (at 1030° C), in addition to the AFM order discussed above, a SC state (above $T_N$=16 K) is induced. Fig. 5 shows the ZFC and FC curves for sample B measured at 5 and 250 Oe. The negative signals in the ZFC curve and in the FC at 5 K, indicate clearly a SC state with an onset at $T_C$= 18 (1) K, a value which was also confirmed by ac susceptibility and resistivity measurements. On the other hand, (i) the a pronounced peak at 11 K in both FC curves and (ii) the merging point of the ZFC and FC branches at $T_N$=16 K (both similar to the asp sample) indicate the AFM ordering of the Mo sublattice. Note that (i) $T_N<T_C$ and (ii) the negative values in the FC curve at 5 Oe and around $T_N$. The shielding fraction (SF) deduced at 5 Oe is ~4% of the -1/4π value, indicating that only a small fraction of the material becomes SC under these conditions. Scanning tunneling microscope measurements performed (STM) at 4.2 K on this sample show that some of the surface grains revealed gapless (nearly Ohmic) tunneling spectra, whereas for most of the grains, clear SC gaps were observed in the tunneling spectra[7]. The gap width exhibited spatial variations in the range 6-7 meV, thus the ratio $2\Delta/k_BT_c$ is around 6.8, within the range observed for various cuprate superconductors [1,10]. The gaps vanished at $T_C$ and therefore are unambiguously associated with the SC state.

The ZFC and FC branches for sample (C) (annealed at 1030° C for two days) are both negative below $T_C$= 23 K (Fig. 6) and the FC branch shows a clear peak at 11 K. The signature of bulk SC is the pronounced field expulsion (the Meissner effect), which appears in the FC curve. Without correcting for diamagnetism, the SF and the Meissner fractions (MF) exceed 12% and 10% respectively. This alone however, cannot be considered as indication for bulk superconductivity, although it would be very difficult to obtain such a signal by an impurity phase in a concentration not detectable with our XRD (Fig. 1). At high applied fields, the M(H) curve at 5 K ( Fig. 6 inset) behaves similarly to that of the asp sample, indicating that the AFM state is not affected by the presence of the SC one. On the other hand, the morphology of sample (C) detected by SEM (Fig. 4b) is somewhat different from that of the non-SC (asp) sample. The grains are approximately spherical, with a typical size of 2-3 μm and are well isolated from each other. Further annealing of sample (C) at 600° C under flowing oxygen (sample D), shifts $T_C$ to 26 K and increases the SF and the MF to 23% and 18% respectively (not shown). The relatively high MF value indicates a weak pinning of the expelled flux lines for this material.

### (c) The high oxygen pressure annealed MoSr$_2$YCu$_2$O$_8$ sample

The ZFC and FC curves for sample (E) (annealed under 92 atm. oxygen) measured at 5 and 500 Oe are shown in Fig 7. $T_C$ = 32 K obtained here, is a much higher than the 18 K for sample B, indicating that $T_C$ depends only on the oxygen concentration. We tend to believe that annealing under higher oxygen pressure (not available in our laboratory) will shift $T_C$ to even higher temperatures. Note (i) the peak around 11 K (at 5 Oe) in both branches and (ii) that the magnetic features are not visible in the 500 Oe curves in which the SC state is the dominant one. The SF and the MF (at 5 K) are ~23% and 18% of the -1/4π value, the same values as for sample D ($T_C$=26 K). We have not corrected for the demagnetization factor, which should be small, since the sample has a bar-shape form and the external field was parallel to the long axis. Therefore we may assume that these SF and MF values represent the bulk properties of the SC state which exists through the entire volume of the material. By the same token, the low SF and MF obtained may be the result of the grain size effect as follows. The penetration depth (λ) for the iso-structure RuSr$_2$GdCu$_2$O$_8$ compound (with the same $T_C$) was evaluated to be around 10 μm[11]. Assuming a similar λ value for the iso-structural MoSr$_2$YCu$_2$O$_8$, causes many grains with a size smaller than λ, not to expel the magnetic field and to reduce the diamagnetic signals in the ZFC and FC curves.

Fig. 8 exhibits the temperature dependence of the real χ'(T) and imaginary χ"(T) ac susceptibility measured at various ac fields ($H_{ac}$). Similar behavior was obtained for Ru-1222[12]. It is readily observed that the broad transition for this granular superconductor occurs via two stages. The intra-grain SC onset ($T_c$=32 K) at which the grains become superconducting, is not affected by ($H_{ac}$). However, at $T_P$ =24 K, due to the weak-Josephson inter-grain coupling, both χ'(T) and χ"(T) are affected dramatically by $H_{ac}$. This behavior is typical for a granular superconductor with weak inter-grain coupling. Generally speaking, below $T_P$ the susceptibility is governed mainly by the weak-links properties, whereas above $T_P$ it is governed by the intra-grain coupling.

With the purpose of acquiring information about the critical current density ($J_C$), we have measured at 5 K the magnetic hysteresis (Fig. 9) of sample (E). Following Bean's approach, $J_C(H) = 30 \Delta M/d$, where $\Delta M$ (in emu/cc) is the difference in M at the same H, and d=2.5 μm (Fig. 4). The $J_C$ is obtained are $J_C = 1.8 *10^5$ and $1.0 *10^5$ A/cm$^2$ at H=0 and 1000 Oe respectively, values, which compare well with $J_C$ obtained in Ru-1222 under the same conditions[12]. The linear field dependence of $J_C$ on a semi-logarithmic scale (with a slope of –0.66) is shown in Fig. 9 (inset).

Above $T_C$, the dc susceptibility curves at 10 kOe measured for all samples have the typical PM shape and adheres to the Curie-Weiss (CW) law: $\chi = \chi_0 + C/(T-\theta)$, where $\chi_0$ is the temperature independent part of $\chi$, C is the Curie constant, and $\theta$ is the CW temperature. Generally speaking, the PM values obtained for all samples are similar. The fits to the CW law yield: C=0.40-0.46 emu K/mol Oe and a negative $\theta$= from -7.6 to -2.2 K, which corresponds to an effective moment $P_{eff}$ =1.79(2)-1.92(2) $\mu_B$. Note that neither the Y ions nor the Puali- paramagnetic $SrMoO_4$ phase contribute to C. Also the roughly temperature independent susceptibility of the Cu ions (1.8-2*10$^{-4}$ emu/mol Oe)[13] does not contribute to C. Therefore, the $P_{eff}$ values obtained, correspond to the Mo ions and are in good agreement with 1.73 $\mu_B$ expected for Mo$^{5+}$ (4d$^1$, S=0.5). Therefore, we argue with high confidence, that the prominent AFM features shown in this paper are related to the Mo$^{5+}$ sublattice.

## Conclusions

Synthetic conditions were optimized for Mo-1212Y, where the Mo$^{5+}$ sublattice is AFM ordered at 16 K. Without a detailed magnetic structure from neutron diffraction studies, it is difficult to comment on the exact AFM structure of this system. The as-prepared sample is not SC. Superconductivity in the Cu-O layers is induced with appropriate annealing. The $T_C$ values as well as the shielding and Meissner fractions, depend strongly on oxygen concentration. The reduced diamagnetic signals can be the result of grain size effect. The highest $T_C$ =32 K was obtained by annealing under high oxygen pressure at 650 C. The AFM ordering, which coexists with SC state through effectively decoupled subsystems, is not affected by the presence or absence of the SC state. In all samples reported here $T_C > T_N$. This rather surprising results raises a question as to why Mo-1212 behave so differently than the Ru-1212 system in which $T_C << T_N$.

**Acknowledgments:** This research was supported by the Israel Academy of Science and Technology and by the Klachky Foundation for Superconductivity.

**References**

[1]. I Felner, U. Asaf, Y. Levi, and O. Millo, Phys. Rev. B 55, R3374 (1997); Physica C 334, 141(2000).

[2]. Y.Y. Xue, B. Lorenz, A. Baikalov, D.H. Cao, Z.G. Li and C.W. Chu, Phys. Rev. B 66, 014503 (2002:: Phys. Rev. B 65, R020511 (2002).

[3]. C. Bernhard, J.L.Tallon, Ch. Niedermayer, Th. Blasius, A. Golnik, B. Btucher, R.K. Kremer, D.R. Noakes, C.E. Stronach and E.J. Ansaldo, Phys. Rev. B 59, 14099 (1999).

[4] I Felner, I. Asaf, U and E. Galstyan. Phys. Rev. B 66, 024503 (2002).

[5] K. Rogacki, B. Dabrowski, O. Chmaissem and J.D. Jorgensen: Phys. Rev. B ,054501(2000)

[6] I. Felner, E. Galstyan Phys. Rev. B, 68, 0645xx(2003).

[7] I. Felner, E. Galstyan, I. Asulin, A. Sharoni, and O. Millo Phys. Rev. Let. (2003) submitted

[8] S.I. Ikeda and N. Shirakawa, Physica *C* 341-348, 785 (2000).

[9] H. Okada, M. Takano and Y. Takeda, Phys. Rev. B 42 (1990) 6831

[10] . H.L. Edwards *et al.*, Phys. Rev. Lett. 69, 2967 (1992).

[11] Y.Y. Xue, B. Lorenz, R.L. Meng, A. Baikalov, and C.W. Chu, Physica 364-365, 251 (2001).

[12] I Felner, E. Galstyan. B. Lorenz, D. Cao, Y.S. Wang Y.Y. Xue, and C.W. Chu, Phys. Rev. B 67, 134506 (2003).

[13] I. Felner, V.P.S. Awana and E. Takayama-Muromachi, Phys. Rev. B 68, (2003) in press.

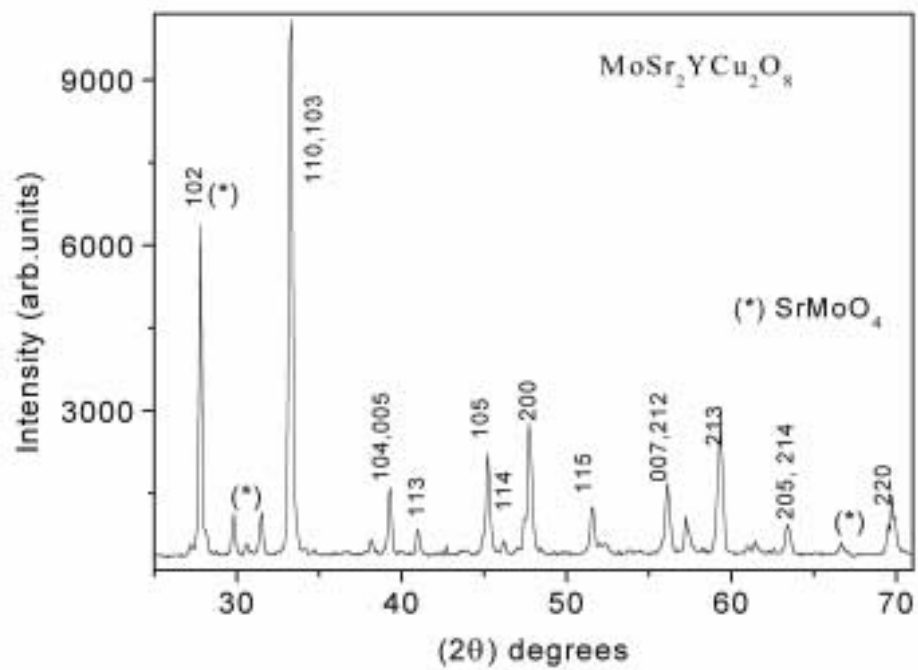

Fig. 1 The XRD patren of $MoSr_2YCu_2O_8$ (sample C)

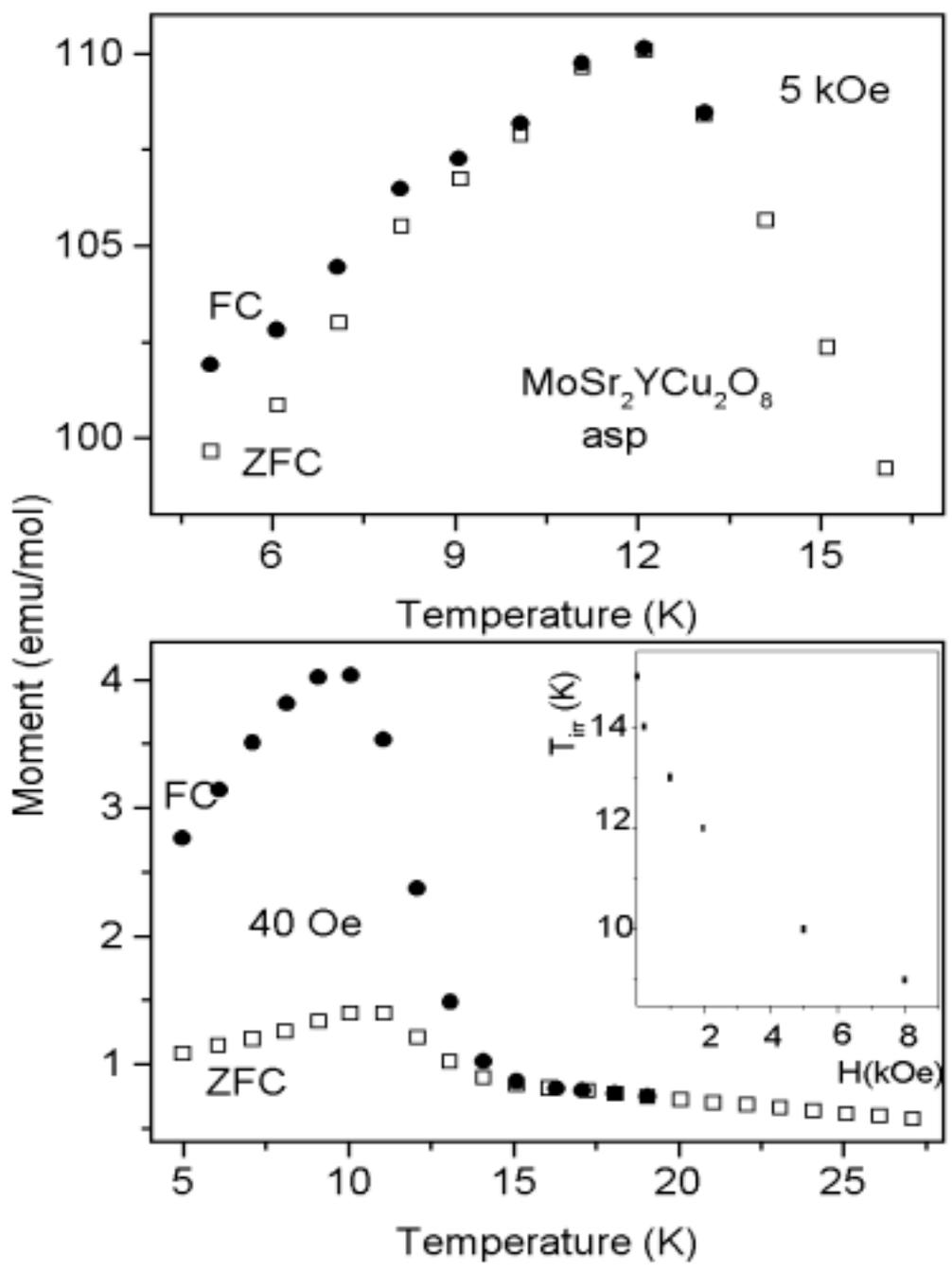

Fig 2. ZFC and FC magnetization curves measured at 40 Oe and 5 kOe of the asp Mo-1212Y sample. The inset shows the the field dependence of the irrevresibility temperature.

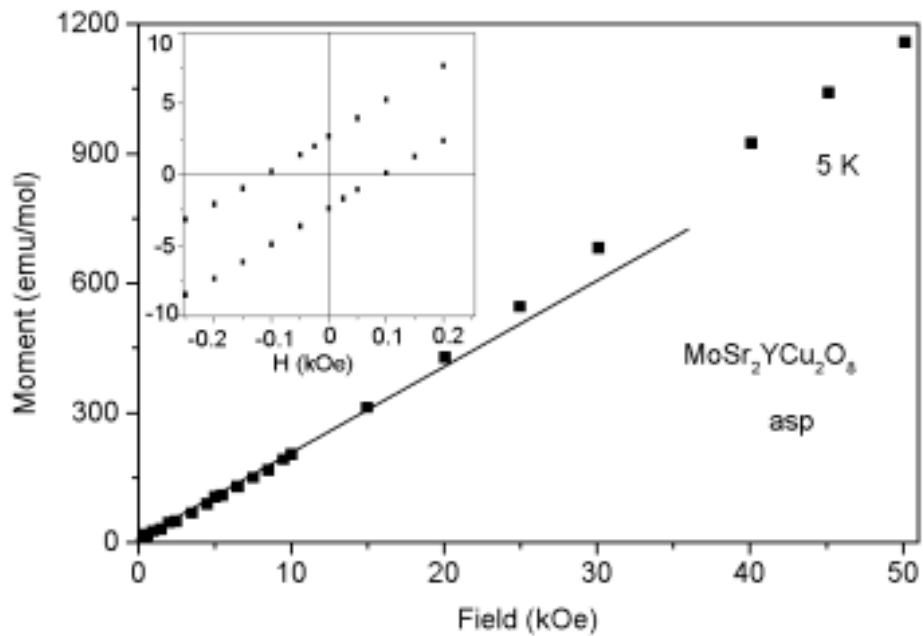

Fig.3. The field dependence of the magnetization curves of the asp MoSr$_2$YCu$_2$O$_8$ sample measured at 5 K. The samll hysteresis loop at low fields is shown in the inset.

**(Not shown due to limited space)**

Fig. 4. SEM pictures of the asp (a) and (b) of sample (C) MoSr$_2$YCu$_2$O$_8$ materials.

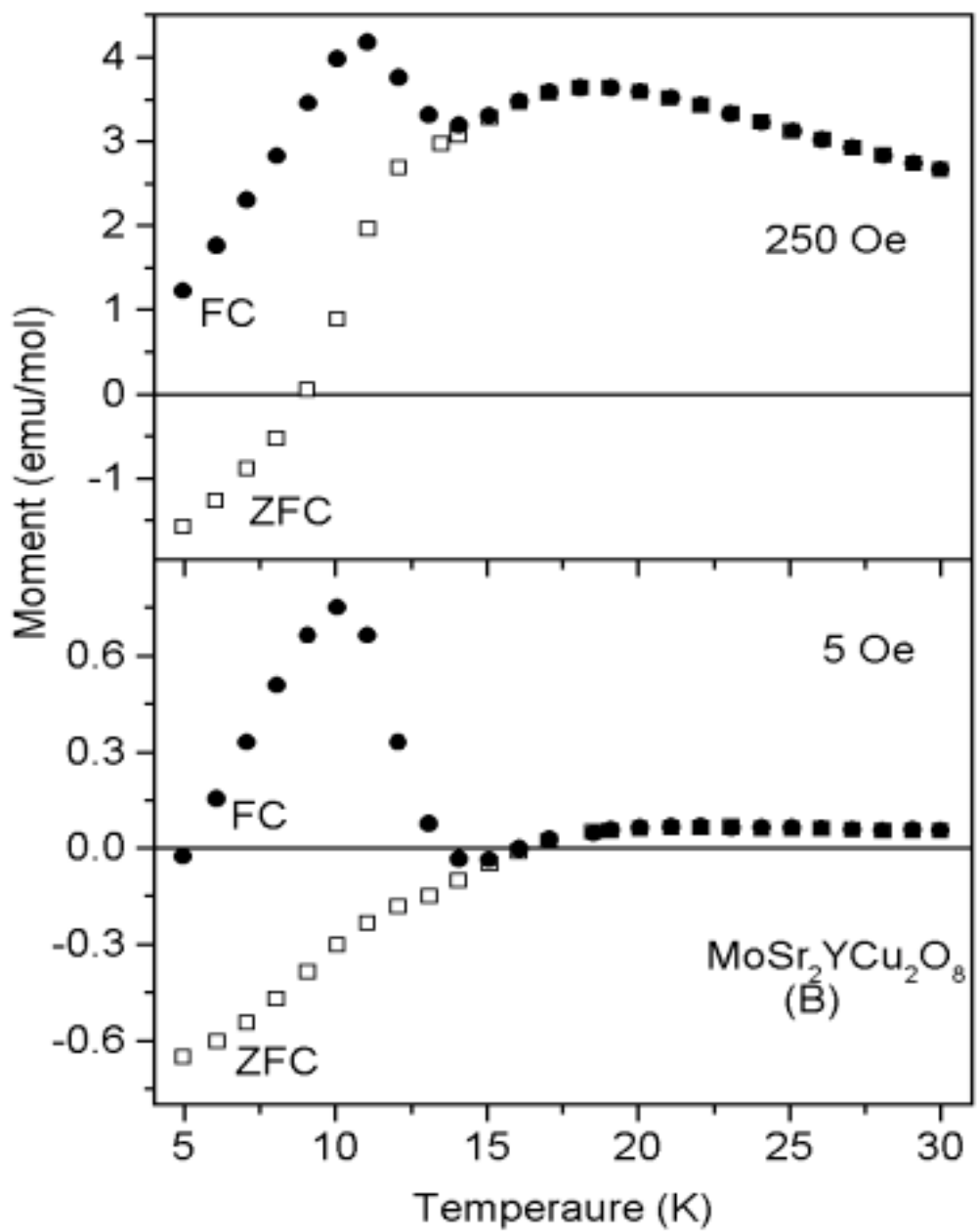

Fig.5. ZFC and FC magnetization curves measured at 5 and 250 Oe for sample (B) ($T_C$=18 K).

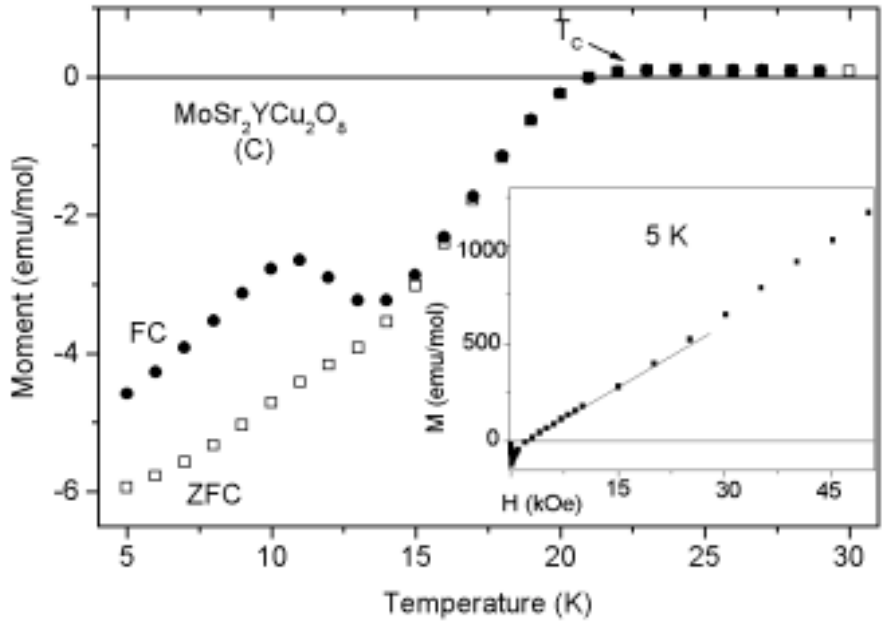

Fig.6. ZFC and FC magnetization curves measured at 5 and 250 Oe and the M(H) plot at 5 K for sample (C) ($T_C$=22 K).

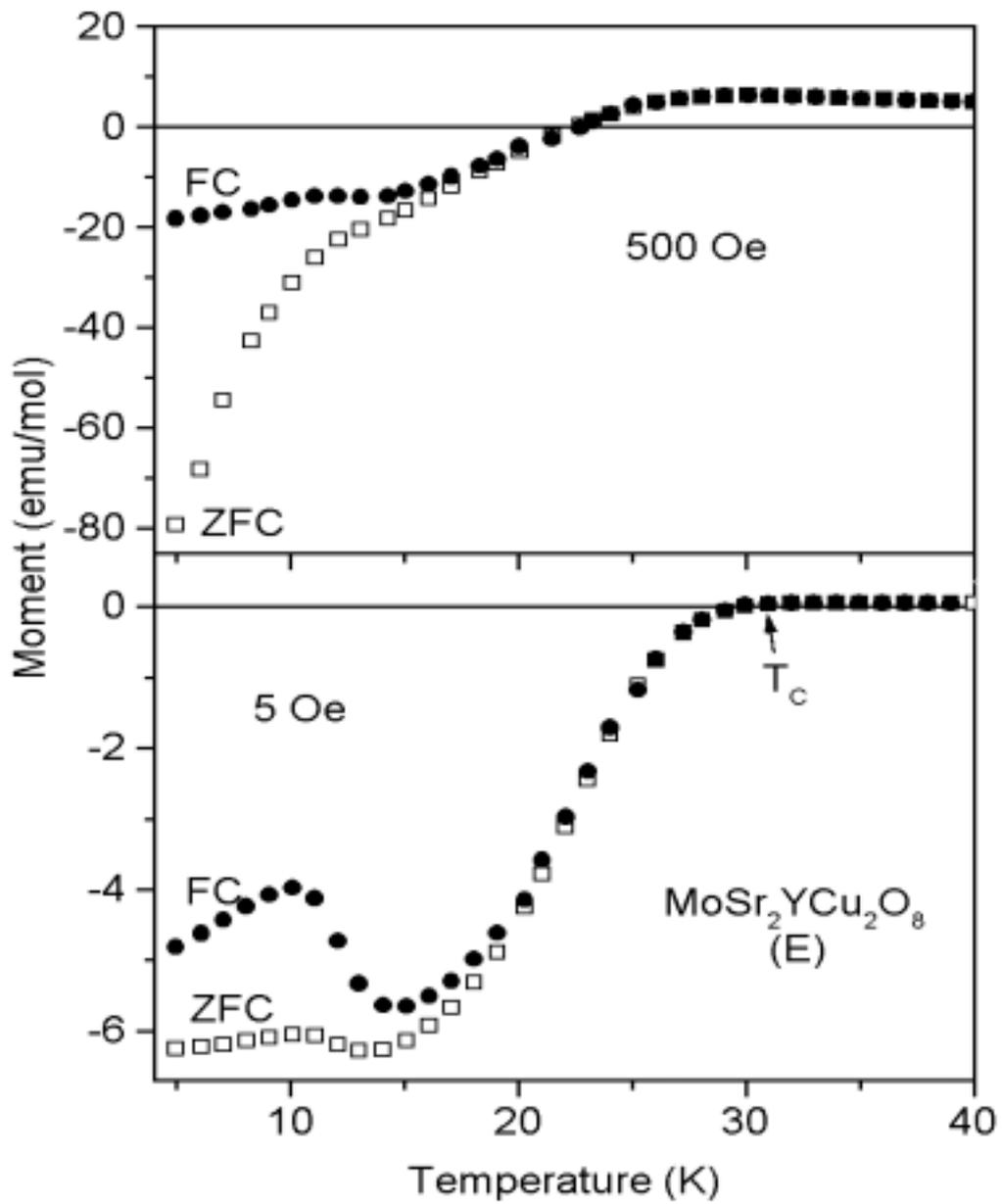

Fig. 7. ZFC and FC magnetization curves measured at 5 and 500 Oe sample (E) ($T_C$=32 K)

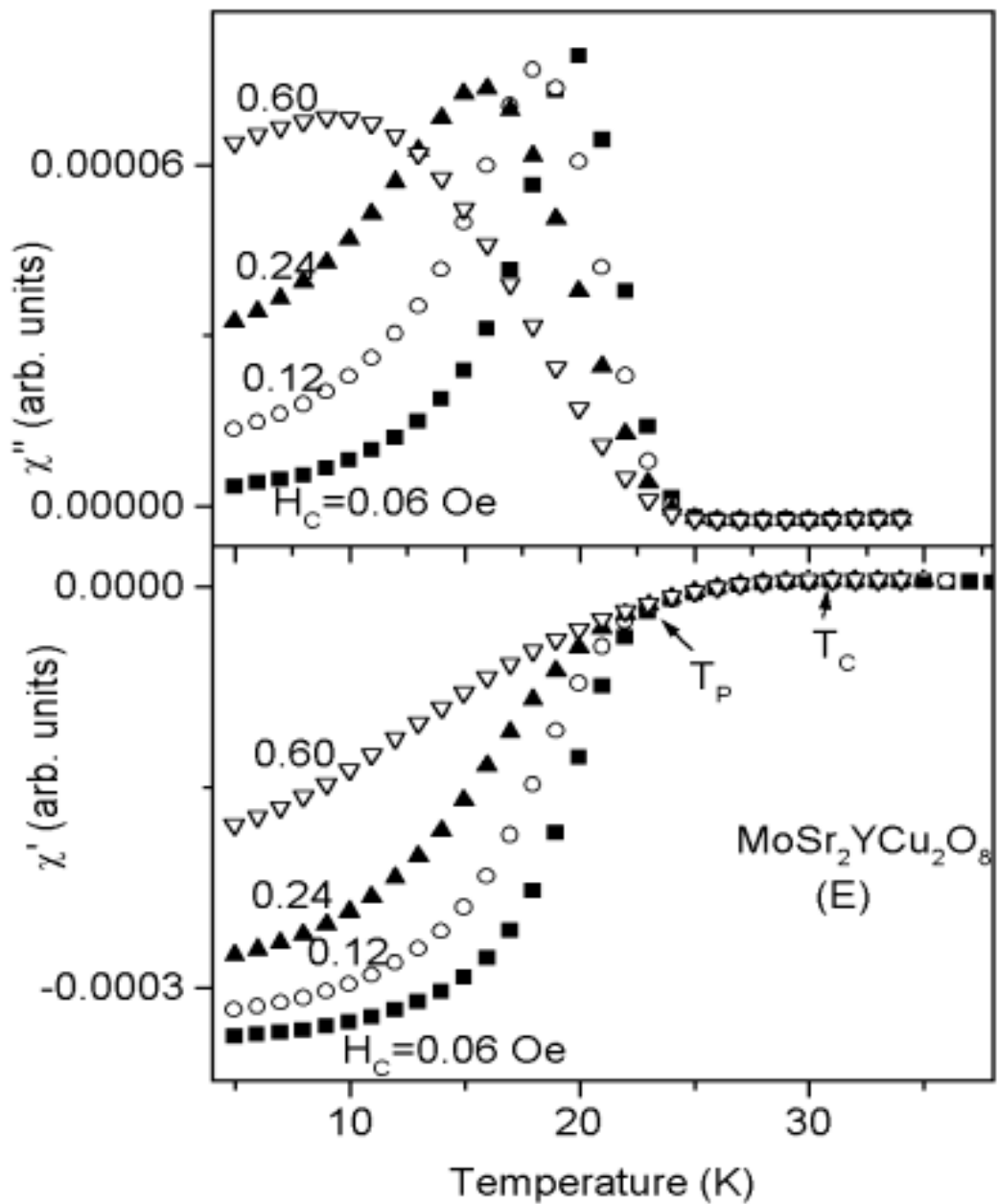

Fig .8 The temperature dependence of the normalized real and imaginary ac susceptibility curves of sample (E).

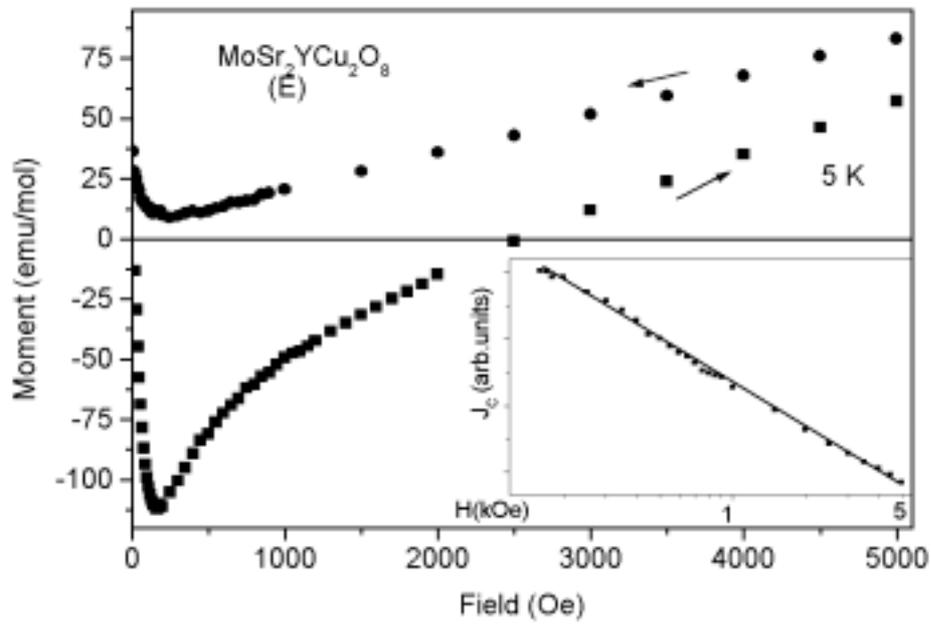

Fig. 9 The SC hysteresis low loop at 5 K and the field dependence of $J_C$ in a semi=logartimic scale, for sample (E).